\documentstyle[12pt]{article}

\newcommand {\nn}    {\nonumber}
\newcommand {\vs}[1]  { \vspace*{#1 cm} }

\newcounter{eq}
\newcounter{sc}

\newcommand {\NP}   {Nucl.Phys.}
\newcommand {\PL}   {Phys.Lett.}

\def\overleftrightarrow#1{\vbox{\ialign{##\crcr
 $\leftrightarrow$\crcr\noalign{\kern-1pt\nointerlineskip}
 $\hfil\displaystyle{#1}\hfil$\crcr}}}

\newlength{\minitwocolumn}
\begin{document}


\begin{flushright}
EDO-EP-28\\
August, 1999\\
\end{flushright}
\vspace{30pt}

\pagestyle{empty}
\baselineskip15pt

\begin{center}
{\large\bf Mass Hierarchy from Many Domain Walls

 \vskip 1mm
}

\vspace{20mm}

Ichiro Oda
          \footnote{
          E-mail address:\ ioda@edogawa-u.ac.jp. 
                  }
\\
\vspace{10mm}
          Edogawa University,
          474 Komaki, Nagareyama City, Chiba 270-0198, JAPAN \\

\end{center}


\vspace{15mm}
\begin{abstract}
We construct a new model with exponential mass hierarchy by starting 
with the Einstein-Hilbert action with the cosmological constant 
in five dimensions plus an action describing many domain walls 
in four dimensions. The model includes many hidden sectors and 
one visible sector, and each four-dimensional domain wall, 
that is, 3-brane, interacts with one another through only 
a gravitational interaction and realizes many universe cosmology
inspired by D-brane perspective.
It is shown that in the present model only even numbers of domain 
walls are allowed to locate in five dimensional space-time
and the validity of Randall-Sundrum scenario, which explains 
mass hierarchy between the Planck mass and the electro-weak scale 
in our world, depends on a relative relation between our world
and hidden worlds. 

\vspace{15mm}

\end{abstract}

\newpage
\pagestyle{plain}
\pagenumbering{arabic}


\rm

Recently, there has been considerable interest and activity in the
higher dimensional theories to solve the hierarchy problem. 
It is well known that in the Standard Model based on the gauge
group $SU(3) \times SU(2) \times U(1)$ there exist two scales, 
those are, the electro-weak scale  $m_{EW} \approx 10^3$ GeV and 
the Planck scale $M_{Pl} \approx 10^{19}$ GeV.
The hierarchy problem is a problem concerning the two mass scales.
The light Higgs mass  $m_{EW} \approx 10^3$ GeV is needed in the
Standard Model, but the quadratic divergences at the loop levels
and the renormalization effects associated with the light Higgs field
cause this electro-weak mass to have the very huge Planck mass $M_{Pl} 
\approx 10^{19}$ GeV, which corresponds to the cut-off scale
of the ultra-violet divergences.
Thus, solving the hierarchy problem is equivalent to explaining 
the vast disparity and quantum stability of the two scales 
without any fine tuning of parameters at each perturbation level.

In recent works, it has been pointed out that the existence of
large extra compact spatial dimensions might give us a solution
for the hierarchy problem \cite{Arkani, Antoni}. 
The key idea is the following: in the simple case, 
the four-dimensional Planck scale $M_{Pl}$ and the 
$(4+n)$-dimensional Planck scale $M_*$ have a 
relation $M_{Pl}^2 = M_*^{n+2} R_c^n$ where $R_c$ is a 
compactification radius. Even if $M_*$ is around the order of
the electro-weak scale,  $M_{Pl}$ can become the Planck scale 
by taking $R_c$ to be large enough.
But against this naive expectation,
it was found that in this scenario the mass hierarchy is merely 
converted into another hierarchy problem between the electro-weak scale 
and the compactification radius.

Subsequently, a new scenario has been proposed for the hierarchy
problem without appealing to large extra compactification 
dimensions \cite{Randall}.
According to this new scenario, the large hierarchy of the Planck scale 
and the electro-weak scale is explained by the nontrivial redshift
factor arisinging from the boundary condition where an extra 
dimension has a structure of an orbifold $S^1/Z_2$. 
The model used in the analysis is obtained by M-theory compactified 
on an $S^1/Z_2$ and a deformed Calabi-Yau threefold 
\cite{Witten, Stelle} and has been examined in detail 
by several groups \cite{Nihei, Kaloper,
Mersini, Verlinde, Kehagias, Wise, Dimo, Lykken}.

The above model \cite{Randall} describes our world as one of 
four-dimensional two slabs in five-dimensional space-time 
while the other slab is regarded as a hidden world 
which interacts with our world by a gravitational interaction.
Although the model is quite of interest from the perspective of
M-theory as mentioned above, it might be possible to construct 
a new model based on the idea of D-branes \cite{Polchinski} 
where an arbitrary number of paralell D-branes are located
at various places in the fifth dimensional direction
in the bulk and interact with one another by a gravitational force. 
Indeed, such a model also appears to be welcome from  
cosmology since in modern cosmology our universe (visible 3-brane)
is conjectured to be not a unique and selected universe but have been 
created with other many universes (hidden 3-branes) simultaneously 
on an equal footing in the big bang era.
Thus it is interesting to construct a new model with one visible
world and many hidden worlds and ask how mass hierarchy 
(and cosmological constant) problem is resolved. 
The purpose of this paper is to pursue this line of thoughts, present 
such a concrete model and then show that the validity of 
a scenario proposed by Randall-Sundrum \cite{Randall} 
explaining mass hierarchy depends on a relative relation between 
our world and hidden worlds.
As a bonus, it turns out that one disadvantage of the previous model,
the necessity for a singular orbifold geometry \cite{Lykken}, will
disappear in the present model since the existence of solutions
of the Einstein equation demands us to choose a smooth manifold $S^1$ 
in place of a singular orbifold geometry $S^1/Z_2$ when there are 
more than two branes in a theory.

We start with the Einstein-Hilbert action with the cosmological
constant in five dimensions plus an action describing $n$ domain 
walls in four dimensions \cite{Randall, Nihei}:
\begin{eqnarray}
S = \frac{1}{2 \kappa^2} \int d^4 x \int_{0}^{2L} dz 
\sqrt{-g} \left(R + 2 \Lambda \right) 
+ \sum_{i=1}^{n} \int_{z=L_i} d^4 x \sqrt{-g_i} {\cal L}_i,
\label{1}
\end{eqnarray}
where the cosmological constant $\Lambda$ is taken to a positive
number,
which implies that the geometry of five-dimensional bulk is 
anti-de Sitter space-time. (Note that we have changed the
sign in front of the cosmological constant compared to a
conventional choice so the positive cosmological constant
corresponds to the anti-de Sitter space-time.)
The fifth dimension $z$ is compact with the size $2L$.
Moreover, $\kappa$ denotes the five-dimensional gravitational
constant with a relation $\kappa^2 = 8 \pi G_N = \frac{8 \pi}{M_*^2}$ 
where $G_N$ and $M_*$ are the five-dimensional Newton constant
and the five-dimensional Planck scale, respectively. 
In this article we follow the standard 
conventions of the textbook of Misner, Thorne and Wheeler \cite{Misner}. 
It is edifying to mention two differences between our starting action 
and that of previous works \cite{Randall}.  
An obvious difference is that we have introduced not two domain walls 
but $n$ many domain walls located at $L_i$ in the $z$ direction.\footnote
{In recent work \cite{Dimo}, many branes are also considered from a 
different context.} (Precisely speaking, all $z=L_i$ are not always
occupied by domain walls. As shown later, the Einstein equation fixes
locations of domain walls completely.)
Another important difference lies in the fact that in the previous M-theory 
model \cite{Randall} the two domain walls are located at the boundaries 
which are the fixed points of an orbifold $S^1/Z_2$, whereas in the
present D-brane model the fifth dimension has a topology of $S^1$
so that various domain walls are not located at the boundary.

Variation of the action (\ref{1}) with respect to the five-dimensional
metric tensor leads to the Einstein equation:
\begin{eqnarray}
\sqrt{-g} \left( R^{MN} - \frac{1}{2} g^{MN} R \right)
= \sqrt{-g} g^{MN} \Lambda 
+ \kappa^2 \sum_{i=1}^{n}  \sqrt{-g_i} g_i^{\mu\nu}
\delta_{\mu}^M \delta_{\nu}^N {\cal L}_i \delta(z - L_i),
\label{2}
\end{eqnarray}
where $M, N, ...$ denote five-dimensional indices, whereas
$\mu, \nu, ...$ do four-dimensional ones. In deriving this 
equation, we have neglected the contributions from the 
Lagrangians ${\cal L}_i$ of domain walls. In other words,
the domain wall actions play a role as sources only providing
the cosmological constant in the bulk.

In order to solve the equation (\ref{2}) in an analytical
way, we require the following metric ansatz
in such a way that the red-shift factor and expanding 
universe are taken into account:
\begin{eqnarray}
ds^2 &=& g_{MN} dx^M dx^N  \nn\\
&=& u(z)^2 \left( -dt^2 + v(t)^2 d\vec{x}^2 \right) + b(z)^2 dz^2,
\label{3}
\end{eqnarray}
where $\vec{x}$ denotes the three-dimensional spatial coordinates.
Under the ansatz (\ref{3}), the Einstein equation (\ref{2}) 
reduces to three combined differential equations:
\begin{eqnarray}
6 \left(\frac{u'}{u} \right)^2 -3 \left(\frac{b}{u} \right)^2
\left[\frac{\ddot v}{v} + \left(\frac{\dot v}{v} \right)^2 \right] 
= b^2 {\Lambda},
\label{4}
\end{eqnarray}
\begin{eqnarray}
3 \left[\frac{u''}{u} + \left(\frac{u'}{u} \right)^2 \right]
-3 \frac{b'}{b} \frac{u'}{u} - 3 \left(\frac{b}{u} \right)^2 
\left(\frac{\dot v}{v} \right)^2
= b^2 \Lambda + \kappa^2 b \sum_{i=1}^{n} {\cal L}_i
\delta(z - L_i),
\label{5}
\end{eqnarray}
\begin{eqnarray}
3 \left[\frac{u''}{u} + \left(\frac{u'}{u} \right)^2 \right]
-3 \frac{b'}{b} \frac{u'}{u} - \left(\frac{b}{u} \right)^2 
\left[ 2 \frac{\ddot v}{v} + \left(\frac{\dot v}{v} \right)^2 \right] 
= b^2 \Lambda + \kappa^2 b \sum_{i=1}^{n} {\cal L}_i
\delta(z - L_i),
\label{6}
\end{eqnarray}
where the prime and the dot denote a differentiation with
respect to $z$ and $t$, respectively.

Now from Eqs. (\ref{5}), (\ref{6}), it is straightforward to derive 
the equation for variable $v$
\begin{eqnarray}
\frac{\ddot v}{v} = \left(\frac{\dot v}{v} \right)^2,
\label{7}
\end{eqnarray}
which can be integrated to
\begin{eqnarray}
v(t)= v(0) e^{H t},
\label{8}
\end{eqnarray}
where $H$ denotes the Hubble constant of expanding universe
\cite{Nihei, Kaloper}. The implications of the present model
to cosmology will be examined in a separate paper \cite{Oda} 
so we shall take $v(t) =1$ from now on in order to focus our
attention on mass hierarchy. Moreover, we set $b(t) = r_c$
where $r_c$ is a constant.

Under the simplified ansatz $v(t) =1$ and $b(t) = r_c$, 
Eqs.(\ref{4}), (\ref{5}) (or (\ref{6})) take rather simple forms
\begin{eqnarray}
\left(\frac{u'}{u} \right)^2  = k^2 r_c^2,
\label{9}
\end{eqnarray}
\begin{eqnarray}
\frac{u''}{u} 
= k^2 r_c^2 + \frac{\kappa^2}{3} r_c 
\sum_{i=1}^{n} {\cal L}_i \delta(z - L_i),
\label{10}
\end{eqnarray}
where we have defined 
\begin{eqnarray}
k = \sqrt{\frac{\Lambda}{6}}.
\label{11}
\end{eqnarray}

Now we wish to find a solution which satisfies the above Einstein
equations (\ref{9}), (\ref{10}). However, it seems to be difficult
to find a general solution, so instead we aim to find some special
solutions closely relating to a model with exponential 
mass hierarchy, whose concrete expression is given by
\begin{eqnarray}
u(z) =  e^{- k r_c f(z)},
\label{12}
\end{eqnarray}
where $f(z)$ is a certain function of variable $z$ to be
determined in what follows.
The most effective and easiest way of determining the function $f(z)$
is to construct it in a direct manner according to the following
procedure.
First, divide the region $0<z<2L$ into $(n-1)$ small pieces
$L_i<z<L_{i+1} (i=1, 2, \cdots, n-1)$.
Second, combine two adjacent pieces into
one pair in order from $L_1=0$ or $L_n=2L$, from which we have
$\frac{n-1}{2}$ distinct pairs. (We assume that $n$ is an odd
number for a while. This issue will be argued later.)  
Third, draw a piecewisely continuous straight line with a slope
$+1$ or $-1$ in each pair by turns.
Finally, since the topology of the fifth dimension is a circle
$S^1$ the two boundaries $z=L_1 \equiv 0$ and $z=L_n \equiv 2L$ 
are identified by setting the periodic boundary condition.
Then except one point mentioned shortly,
it is straightforward to show that the line obtained in this
way satisfies the Einstein equations (\ref{9}), (\ref{10})
by using the Fourier series expansion.

Here one would like to discuss an important subtlety associated with
the definition of $\varepsilon(0)$. 
In showing that Eq.(\ref{12}) with $f(z)$ obtained in the above
procedure satisfies the Einstein equations (\ref{9}), (\ref{10})
one encounters the step function $\varepsilon(x)$ defined as
\begin{eqnarray}
\varepsilon(x) = \frac{|x|}{x} 
= \cases{+1, &for $x > 0$ \cr  
         -1, &for $x < 0$ \cr}.
\label{13}
\end{eqnarray}
At some $L_i$ on which domain walls sit, 
we have an ambiguous quantity $\varepsilon(0)$, 
for which we usually define as $\varepsilon(0)=0$ or $\varepsilon(0)=1$. 
The validity of the Einstein equation now requires
us to choose $\varepsilon(0)=1$. Thus at first sight we do not
meet any inconsistency, but various results should not depend on
such an ambiguous quantity.
Indeed, in the situation at hand this quantity has a close 
connection with a physical fact, namely, the existence of singularities 
at the location of domain walls, so a careful treatment is needed. 
In this article, in order to avoid  an ambiguous quantity $\varepsilon(0)$, 
we wish to regularize our model in a such way 
that we introduce an infinitely small thickness $\Delta$ of domain walls 
and take a limit $\Delta \rightarrow 0$ after all calculations. 
Under this regularization, any ambiguity never occur. 

Let us present two interesting solutions. These solutions
describe even domain walls standing along $S^1$ at some intervals
in five dimensional space-time.
One solution is given by
\begin{eqnarray}
f(z) &=& |z| +  \sum_{i=1}^{\frac{n-1}{2}} (-1)^i |z - L_{2i}| - L, 
\nn\\
f'(z) &=& \sum_{i=1}^{\frac{n-1}{2}} (-1)^i \varepsilon(z - L_{2i}) + 1, 
\nn\\
f''(z) &=& 2 \sum_{i=1}^{\frac{n-1}{2}} (-1)^i \delta(z - L_{2i}),
\label{14}
\end{eqnarray}
where $L_i$ satisfies the relations
\begin{eqnarray}
L_{2i} = \frac{L_{2i-1} + L_{2i+1}}{2}, \ L_1 = 0, \ L_n = 2L,
\label{15}
\end{eqnarray}
with $i = 1, 2, \cdots, \frac{n-1}{2}$. Moreover, ${\cal L}_i \
(i = 1, 2, \cdots, \frac{n-1}{2})$ must satisfy the relations
\begin{eqnarray}
{\cal L}_{2i} = (-1)^{i+1} \frac{6k}{\kappa^2 r_c}.
\label{16}
\end{eqnarray}
This solution expresses $\frac{n-1}{2}$ even domain walls locating 
at $L_{2i}$. In this paper,
we would like to take account of one visible 3-brane and many
hidden 3-branes more than or equal to 1 so $n$ runs over
$5, 9, 13, \cdots$.

Another interesting solution takes the form
\begin{eqnarray}
f(z) &=& \sum_{i=2}^{n-1} (-1)^{i+1} |z - L_i| + L, 
\nn\\
f'(z) &=& \sum_{i=1}^{n-1} (-1)^{i+1} \varepsilon(z - L_i) - 1, 
\nn\\
f''(z) &=& 2 \sum_{i=1}^{n-1} (-1)^{i+1} \delta(z - L_i),
\label{17}
\end{eqnarray}
where $L_i$ also satisfies Eq.(\ref{15}). 
This time, ${\cal L}_i \ (i = 1, 2, \cdots, \frac{n-1}{2})$ must satisfy 
the relations
\begin{eqnarray}
{\cal L}_1 &=& {\cal L}_n = - \frac{3k}{\kappa^2 r_c}, \nn\\
{\cal L}_{2i} &=& - {\cal L}_{2j+1} = \frac{6k}{\kappa^2 r_c},
\label{18}
\end{eqnarray}
where $0 <j < i$.
Now the number of domain walls is an even number $n-1$ 
($n$ takes values $3, 5, 7, \cdots$), and the domain walls 
are located at $L_i \ (i=1, 2, \cdots, n-1)$.
At this stage, one might ask whether it is possible to find
a solution with odd domain walls or not. The answer does not seem to 
be affirmative. The reason is that as in an orbifold geometry $S^1/Z_2$
\cite{Randall}, the solutions obtained in this paper have a characteristic
feature that number of positive energy branes equals to number of 
negative energy branes, so the resulting solutions always include
even domain walls.

Now let us turn to our main question. "How is mass hierarchy 
problem resolved in our many universe model?" We will
see that the answer is quite of interest and shows a
peculiar feature of many universe model.
First of all, let us consider the first model (\ref{14}).
In this model, domain walls are located at $z=L_{2i} \
(i=1, 2, \cdots, \frac{n-1}{2})$.
Let us suppose that a visible 3-brane (our universe) is 
located at a certain $z=L_{2i}$, whereas the remaining 3-branes 
are hidden sectors (the other universes).
We wish to measure mass scale in our world placed at $z=L_{2i}$
by mass scale in a hidden world placed at $z=L_{2j} (i \neq j)$, 
to which we assume that the Planck mass is assigned. 
According to formulas given in \cite{Randall}, we can in general 
evaluate mass scale in our world by using Eqs.(\ref{12}), (\ref{14}):
\begin{eqnarray}
m(L_{2i}) &=& \frac{u(L_{2i})}{u(L_{2j})} m(L_{2j}) \nn\\
&=& m(L_{2j}) \exp\left\{{- k r_c \left[L_{2i} - L_{2j} + 
\sum_{l=1}^{\frac{n-1}{2}} (-1)^l \left(|L_{2i} - L_{2l}|
- |L_{2j} - L_{2l}| \right) \right]}\right\}.
\label{19}
\end{eqnarray}

To compare the result (\ref{19}) to that of \cite{Randall},
let us first recall the result of \cite{Randall} and then
see the implications of our result in the two specific cases.
In the setup of Ref.\cite{Randall}, there are two 3-branes with 
opposite sign of potential energy. For a 3-brane with positive
potential energy, a natural scale for mass is assumed to be
of order the Planck mass. If the Standard Model (our universe) is
placed on another 3-brane with negative potential energy, the
graviton amplitude is exponentially suppressed and exponential
mass hierarchy is generated. 
To have this fact in mind, let us see what happens in our
model. From the action (\ref{1}) and (\ref{16}), it turns out that
3-branes located at $z=L_{4i-2}$ have negative brane energy while
3-branes at $z=L_{4i}$ have positive brane energy. Thus,
for instance, let us calculate mass scale at $z=L_2$ from
mass scale at $z=L_4$, for which we assume that the Planck
mass is assigned. Note that this situation is similar to
that of \cite{Randall} as mentioned above. A simple calculation
yields
\begin{eqnarray}
m(L_2) = m(L_4) e^{- k r_c (L_4 - L_2)}.
\label{20}
\end{eqnarray}
This result means that exponential mass hierarchy also occurs 
in the present model as in \cite{Randall}. Note that this result is 
universal in the sense that the essential behavior of the result remains
unchanged as long as our 3-brane has negative brane energy and the hidden
3-brane has positive brane energy even if a factor $L_4 - L_2$ in
the exponential changes depending on which 3-branes we consider. 
An interesting feature of our model is the existence of many
branes with negative and positive brane energy, so we are now led
to ask how the result about mass hierarchy is modified 
when we take account of two 3-branes with same sign of brane
energy. To be specific, let us consider a situation where we attempt to
measure mass scale at $z=L_2$ from mass scale at $z=L_6$,
of which both the 3-branes have negative brane energy. 
In this case the general result (\ref{19}) reduces to
\begin{eqnarray}
m(L_2) = m(L_6) e^{- \frac{1}{2} k r_c (L_3 + L_5 - L_7)},
\label{21}
\end{eqnarray}
which implies that there is also exponential mass hierarchy
if $L_3 + L_5 > L_7$. But if this inequality does not hold,
then the Planck mass scale allocated in hidden sector is exponentially 
enhanced to the larger mass scale in our world. It is easy
to check that the essential feature of this conclusion is
also universal as far as two 3-branes with same sign of brane
energy are concerned.

Next, let us focus on the second model (\ref{17}).
In this model, domain walls are located at $z=L_i \
(i=1, 2, \cdots, n-1)$. Again, a simple calculation gives us
\begin{eqnarray}
m(L_i) &=& \frac{u(L_i)}{u(L_j)} m(L_j) \nn\\
&=& m(L_j) \exp\left\{{- k r_c \sum_{l=2}^{n-1} (-1)^{l+1} 
\left(|L_i - L_l| - |L_j - L_l| \right) }\right\}.
\label{22}
\end{eqnarray}
{}From the action (\ref{1}) and (\ref{18}), we see that 3-branes
located at $z=L_{2i}$ have negative brane energy while
3-branes at $z=L_{2i-1}$ have positive brane energy. Thus,
as in the first model, for instance, let us calculate mass scale 
at $z=L_2$ from mass scale at $z=L_1=0$, for which we take the 
Planck mass. The result is of the form
\begin{eqnarray}
m(L_2) = m(0) e^{- k r_c L_2}.
\label{23}
\end{eqnarray}
Indeed, this result is the same as in \cite{Randall}. (Recall that
for comparison we have to set $L_2$ to $\pi$.) An interesting thing 
here is that this holds true even if there are many hidden sectors
and the topology is a circle $S^1$ instead of an orbifold $S^1/Z_2$.
Therefore, again we have exponential mass hierarchy even in the 
second model. Moreover, it is straightforward to show that even 
if we take account of two 3-branes with same sign of brane
energy we have also exponential mass hierarchy if a suitable inequality 
holds like the first model.
Finally, let us consider mass scale of branes at $z=L_{2i-1}$. 
Since we have an equation $u(L_{2i-1})=1$, the branes at these locations
have special feature. Namely, if we attempt to evaluate mass scale 
at $z=L_{2i-1}$ from mass scale at $z=L_{2j-1}$ with $i \neq j$,
there is no mass hierarchy, in other words, $m(L_{2i-1})=m(L_{2j-1})$.
However, this is an artifact of the model. In fact, we can modify
the value of $u(L_{2i-1})$ from one to a non-zero constant
without violating the validity of the Einstein equation, so in such
a situation we have exponential mass hierarcy as before.

To summarize, we have investigated a possibility of
constructing a new model with an exponential mass hierarchy
whose existence is inspired by the perspective of
D-brane theory and many universe cosmology.
In our model, the fifth dimension has a topology $S^1$
rather than a singular orbifold $S^1/Z_2$, which is
one of advantages in our model. 
It was shown that even in the present model, we have the 
exponential mass hierarchy under an appropriate condition.
However, there are many universes in the present model,
perhaps our universe and the other many hidden universes,
so there is $\it{a \ priori}$ no way of determining which
universe our universe is among many universes. Hence,
the correct interpretation of our results is the following:
mass scale in our universe depends on a relative distance relation
along the fifth dimension between our universe and hidden 
universes.
Maybe once God created many universes and gave the Planck scale
to one (or some) universe(s) in the beginning of the big bang,
mass scales in various universes are fixed by a relative relation
among universes. According to this new scenario, each universe
should have $\it{not}$ more than one $\it{but}$ exactly one
low (or high) energy scale, and our world happens to have
taken the electro-weak scale as such a low energy scale.
Although we have so far presented only two types of solutions
which satisfy the Einstein equation and have simple and
manageable forms, we have also examined the other solutions
in some detail \cite{Oda}. The conclusion is almost the same
as in the present cases,
so we think that the two solutions account for the essential
features in our theory.
One disadvantage of our model as well as the Randall-Sundrum
original model \cite{Randall} is the existence of 3-branes
with negative potential energy. Recently, an alternative setup
has been put forward where only positive energy objects are
taken into account \cite{Lykken}. 
We wish to investigate a possibility of
generalizing our model to such a direction in future.

\vs 1
\begin{flushleft}
{\bf Acknowledgement}
\end{flushleft}
We are indebted to M. Tonin for stimulating
discussions and continuous encouragement. 
We wish to thank Dipartimento Di Fisica, "Galileo Galilei", 
Universita Degli Studi Di Padova, for a kind hospitality, where 
most of parts of this work have been done.

\vs 1

\end{document}